\documentclass{egpubl}
\usepackage{pg2023}

\SpecialIssuePaper         % uncomment for final version of Computer Graphics Forum, special issue
\CGFccbync
\usepackage[T1]{fontenc}
\usepackage{dfadobe}  

\usepackage{cite}  % comment out for biblatex with backend=biber

\BibtexOrBiblatex
\electronicVersion
\PrintedOrElectronic
\ifpdf \usepackage[pdftex]{graphicx} \pdfcompresslevel=9
\else \usepackage[dvips]{graphicx} \fi

\usepackage{egweblnk}
\usepackage[title, toc, page]{appendix}
\usepackage{amstext}
\usepackage{amsfonts}
\usepackage{amsmath}
\usepackage{amssymb}
\usepackage{kotex}

\usepackage{booktabs} 
\usepackage{subcaption}
\usepackage{multicol}
\usepackage{multirow}

\usepackage[ruled]{algorithm2e} 

\SetAlFnt{\small}
\SetAlCapFnt{\small}
\SetAlCapNameFnt{\small}
\SetAlCapHSkip{0pt}

\usepackage{enumitem}
\usepackage{color}
\definecolor{blue}{rgb}{0,0,0.8}
\definecolor{green}{rgb}{0,0.6,0}
\definecolor{red}{rgb}{0.7,0,0}

\usepackage[normalem]{ulem}

\def\vec#1{\mathbf{#1}}

\newcommand{\vg} {\vec{g}}

\newcommand{\vp} {\vec{p}}

\newcommand{\vx} {\vec{x}}

\title{MOVIN: Real-time Motion Capture using a Single LiDAR}

\author[D-K. Jang \& DS. Yang \& D-Y. Jang \& B. Choi \& T. Jin \& S-H. Lee]
{\parbox{\textwidth}
{\centering Deok-Kyeong Jang\thanks{Equal contribution}$^{1,2}$\orcid{0000-0002-7567-4339},
Dongseok Yang$^\dag$$^{1,2}$\orcid{0000-0002-4696-3465},
Deok-Yun Jang$^\dag$$^{1,3}$\orcid{0009-0006-1923-2540},
Byeoli Choi$^\dag$$^{1,2}$\orcid{0000-0003-2347-149X},
Taeil Jin$^{2}$\orcid{0009-0006-6975-9930} and
Sung-Hee Lee\thanks{Corresponding author}$^{2}$\orcid{0000-0001-6604-4709}
}
        \\
{\parbox{\textwidth}{\centering $^1$MOVIN Inc.\\
         $^2$Korea Advanced Institute of Science and Technology (KAIST) \\
       $^3$Gwangju Institute of Science and Technology (GIST)
       }
}
}

\begin{document}

\teaser{
 \captionsetup{labelfont=bf,textfont=it}
 \includegraphics[width=\linewidth]{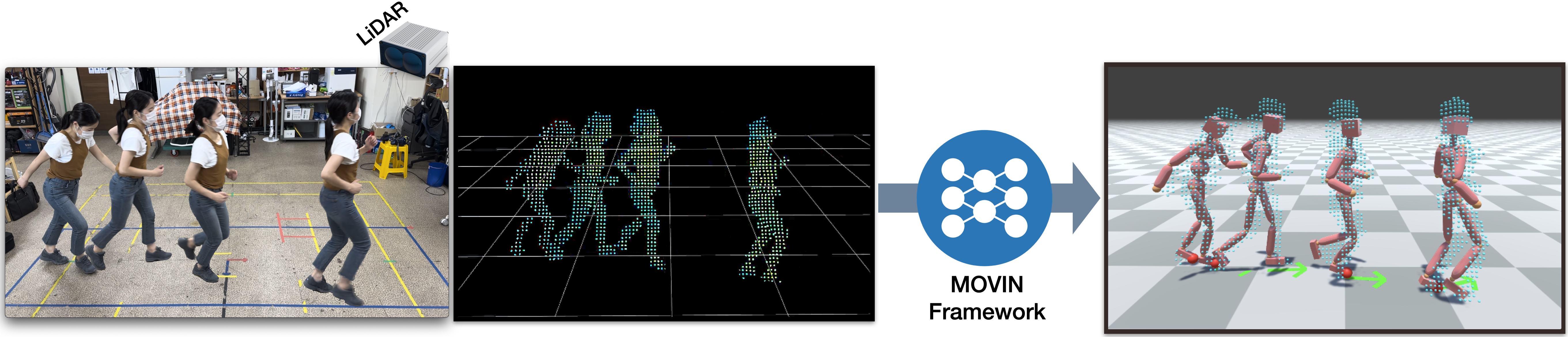}
 \centering
  \caption{Our MOVIN framework enables real-time full-body motion capture with global translation from 3D LiDAR point cloud.}
\label{fig:teaser}
}

\maketitle
%-------------------------------------------------------------------------
\begin{abstract}
Recent advancements in technology have brought forth new forms of interactive applications, such as the social metaverse, where end users interact with each other through their virtual avatars. In such applications, precise full-body tracking is essential for an immersive experience and a sense of embodiment with the virtual avatar. However, current motion capture systems are not easily accessible to end users due to their high cost, the requirement for special skills to operate them, or the discomfort associated with wearable devices.
In this paper, we present MOVIN, the data-driven generative method for real-time motion capture with global tracking, using a single LiDAR sensor. Our autoregressive conditional variational autoencoder (CVAE) model learns the distribution of pose variations conditioned on the given 3D point cloud from LiDAR.
As a central factor for high-accuracy motion capture, we propose a novel feature encoder to learn the correlation between the historical 3D point cloud data and global, local pose features, resulting in effective learning of the pose prior. Global pose features include root translation, rotation, and foot contacts, while local features comprise joint positions and rotations.
Subsequently, a pose generator takes into account the sampled latent variable along with the features from the previous frame to generate a plausible current pose.
Our framework accurately predicts the performer's 3D global information and local joint details while effectively considering temporally coherent movements across frames. We demonstrate the effectiveness of our architecture through quantitative and qualitative evaluations, comparing it against state-of-the-art methods. Additionally, we implement a real-time application to showcase our method in real-world scenarios. MOVIN dataset is available at \url{https://movin3d.github.io/movin_pg2023/}.
%-------------------------------------------------------------------------

%The tool at \url{http://dl.acm.org/ccs.cfm} can be used to generate
% CCS codes.


\end{abstract}
%-------------------------------------------------------------------------

%%%%%%%%%%%%%%%%%%%%%%%%%%%%%%%%%%%%%%%%%%%%%%%%%%%%%%%%%%%%%%%%%%%%%%%%%%%%%%
\section{Introduction}
\label{sec:introduction}
%%%%%%%%%%%%%%%%%%%%%%%%%%%%%%%%%%%%%%%%%%%%%%%%%%%%%%%%%%%%%%%%%%%%%%%%%%%%%%

With the increasing demand for immersive and interactive experiences in the fields of filming, animation, and the metaverse, real-time motion capture has become an essential technology for animating virtual characters to realize interactions with the virtual environment and between the users. However, state-of-the-art motion capture technologies, including optical and inertial sensors, are hardly affordable for general users for their price and inconvenience.

To address the limitations, researchers focused on utilizing commonly accessible sensors, such as cameras and VR tracking devices, to achieve high-quality motion capture in real-time. Recent deep learning approaches robustly predict the full-body pose, leveraging single RGB video stream \cite{li2022cliff}, sparse sets of IMU sensors \cite{huang2018deep}, or sparse VR tracker configurations \cite{yang2021lobstr}. Despite the promising results, there still remains a significant gap that requires improvement; 2D image-based methods suffer from inaccurate global translation, and sparse trackers provide only under-determined constraints that cannot disambiguate different poses with the same tracker configuration. Furthermore, these sensors often face inherent limitations including magnetic interference from surrounding electronics and optical occlusion in complex indoor environments, which can degrade the accuracy and reliability of the captured motion data.

The recent success of LiDAR sensors in object and human detection for autonomous driving \cite{zheng2021se} demonstrates the potential of LiDAR to significantly improve the performance and usability of current motion capture technologies. Unlike traditional 2D camera systems, LiDAR sensors can provide reliable and precise 3D positions of the tracking target in the form of the point cloud. Moreover, this LiDAR-generated point cloud can provide full-body information about the subject, which is not available for sparse configuration of wearable trackers. Pioneering works already proved that LiDAR sensors can assist existing motion capture technologies and increase accuracy for long-range human pose detection \cite{li2022lidarcap}. Furthermore, as LiDAR technology has found applications across various industrial sectors, including security and smart cities, increasing demand has prompted mass production, lowering LiDAR prices.

This paper introduces MOVIN, a novel framework for real-time motion capture using a single LiDAR sensor, as illustrated in Fig.~\ref{fig:teaser}. To the best of our knowledge, our framework is the first LiDAR-based real-time full-body motion capture with global translation.

Our model employs an autoregressive conditional variational autoencoder (CVAE) architecture to establish the relationship between the input point cloud and the output full-body motion, considering the previous output motion. The encoder component, based on the Transformer architecture, maps encoded features to a multivariate Gaussian distribution. Meanwhile, the decoder component follows a Mixture-of-Expert architecture, generating output features by sampling from the distribution while incorporating condition features.

To address the distinct characteristics of the input and conditions, we have carefully designed input/output and feature embedding/expanding modules. The input 3D point cloud integrates current and subsampled data from the past 1-second interval to ensure temporal coherence of the output motion. We process the condition, which represents the output of the previous frame, separately for local and global pose features. The local pose feature of joint local transformations is processed via a skeleton-aware Graph Convolutional Network that preserves inherent body part structure. The global pose feature, which includes root position, rotation, and foot contacts, is handled using Multilayer Perceptrons. Such designs enable our framework to effectively represent and integrate diverse input sources of 3D point cloud, skeletal poses, and global translation and contacts.

For training and evaluation of our framework, we collected a precisely synchronized dataset comprising LiDAR point cloud and optical motion capture data. The dataset involved $10$ subjects with varied body shapes and motion styles, engaging in a wide range of action categories of in-place movements and locomotion. 

To validate the effectiveness of our method, we conducted comprehensive testing on unseen subjects, performing not only daily activities but also challenging motions such as lunging, sitting on the floor, and squatting. Furthermore, we highlight the practicality of our framework by showcasing its real-time application leveraging a single LiDAR sensor, implemented with a commercial game engine.

In summary, this paper presents the following main contributions:
\begin{itemize}[leftmargin=*, itemsep=5.0pt]

\item The real-time full-body motion capture framework based on a single LiDAR, incorporating global translation tracking.

\item A novel design for feature encoding and decoding from different input sources, utilizing an autoregressive conditional variational autoencoder (CVAE) architecture to generate full-body poses from 3D point cloud data.

\item A high-quality dataset featuring diverse subjects, containing synchronized LiDAR point cloud and optical motion capture data for a wide range of actions.

\end{itemize}
%%%%%%%%%%%%%%%%%%%%%%%%%%%%%%%%%%%%%%%%%%%%%%%%%%%%%%%%%%%%%%%%%%%%%%%%%%%%%%
\section{Related Work}
\label{sec:related_work}
%%%%%%%%%%%%%%%%%%%%%%%%%%%%%%%%%%%%%%%%%%%%%%%%%%%%%%%%%%%%%%%%%%%%%%%%%%%%%%
\subsection{Motion Capture}
High-quality motion capture techniques using optical markers  \cite{optitrack-link, vicon-link, vlas2007mocap} and inertial measurement units (IMUs) \cite{xsens-link} have emerged as leading solutions in the industry, offering precise and reliable data for human motion analysis and character animation. For end-users, Vive trackers \cite{vive-link} offers a cost-effective solution. However, current technologies require a large number of markers or sensors on the body and a time-consuming setup process. Therefore, researchers explored alternatives with a sparse setup of IMUs \cite{von2017sparse, yi2021transpose, jiang2022transformer, yi2022physical} and trackers \cite{aliakbarian2022flag, jiang2022avatarposer, winkler2022questsim}. Despite their promising results, these methods still have limitations in the tracking accuracy and coverage of motion categories.

Markerless motion capture techniques have been extensively explored \cite{breg1998maps, agu2008multi, hol2012multi} to enhance the accessibility of motion capture technology, by reducing the cost and improving usability. While multi-view camera algorithms \cite{Amin2013MultiviewPS, Bure2013multiview, Dong2022Multiview} have achieved higher accuracy, they often require laborious camera system calibration. Mono-camera approaches with optimization techniques \cite{bogo2016smpl, kolotouros2019convolutional} and neural networks \cite{pavlakos2017coarsetofine, wei2022capturing, huang2022neural} lack depth information and struggle to track global translations. Despite offering an additional depth channel, RGBD-based solutions \cite{Baak2011depth, VNect_SIGGRAPH2017, Ying2021RgbD} are hindered by limited camera resolution and a field of view (FOV), which makes them impractical for product-level applications.

\subsection{LiDAR-based 3D Human Pose Estimation}
Recent advancements in 3D human pose estimation have seen the emergence of image-based methods like VIBE \cite{kocabas2020vibe} and MotionBERT \cite{zhu2022motionbert}. These methods follow a two-stage process: extracting 2D keypoints and fitting the SMPL model \cite{loper2015smpl} to estimate 3D keypoints. However, relying on 2D fitting poses limitations that compromise the accuracy of 3D keypoints. Additionally, the absence of depth information presents challenges for accurate global tracking.

To tackle these challenges, researchers have explored the integration of LiDAR sensors in 3D human pose estimation. LiDARs offer precise depth measurements, making them well-suited for large-scale environmental measurements for autonomous driving scenes \cite{lang2019pointpillars, shi2020pv}. Recent studies have delved into utilizing LiDAR for capturing detailed 3D human poses \cite{ren2023lidar}. Moreover, sensor fusion approaches combining LiDAR and cameras have been proposed \cite{cong2022weakly, ren2023lidar} to leverage the complementary strengths of these sensors. However, these methods primarily focus on scene-level tasks like human detection and segmentation, rather than capturing skeletal motions with precise global translation.

\subsection{Neural Generative Models for Motion Synthesis}
Motion synthesis has been a prominent area of research, aimed at generating high-quality motion with minimal effort. Initial studies utilized probabilistic methods including principal component analysis (PCA)\cite{safonova04tog, chai05siggraph,
liu06pca}, Gaussian mixture models (GMMs) \cite{min12gmm}, and Gaussian processes \cite{grochow04, wang08gp, levine12gp}.

Generative neural networks have recently gained substantial attention due to their impressive results in character motion synthesis. Multiple methods adopted Generative Adversarial Networks (GANs~\cite{goodfellow14gan}) 
for speech-to-gesture synthesis \cite{ferstl19gesture}, motion control \cite{wang19gans},
and generation from a single motion clip \cite{li22ganimator}. Variational Autoencoder (VAE) is another commonly used architecture that enables random sampling from a specified distribution. Furthermore, conditional VAE(CVAE)~\cite{sohn2015cvae} based methods use constraints such as motion history \cite{ling20character}, motion categories \cite{petrovich21cvae}, and speech \cite{lee19neurips, li20vae} for generation. 
Henter~et~al.~\shortcite{henter20moglow} proposed utilizing normalizing flow for motion generation, enabling efficient training with exact maximum likelihood. Aliakbarian~et~al. \cite{aliakbarian2022flag} extended the work with an additional latent region approximator model. Inspired by the recent accomplishments of diffusion models in computer vision research, Tevet~et~al. \shortcite{tevet22mdm} and Zhang~et~al. \shortcite{zhang22motiondiffuse} proposed language-driven motion synthesis techniques, while Tseng~et~al. \shortcite{tseng22edge} focused on synthesizing dance motion from music.
%%%%%%%%%%%%%%%%%%%%%%%%%%%%%%%%%%%%%%%%%%%%%%%%%%%%%%%%%%%%%%%%%%%%%%%%%%%%%%
\section{MOVIN Dataset}
\label{sec:data_construction}
%%%%%%%%%%%%%%%%%%%%%%%%%%%%%%%%%%%%%%%%%%%%%%%%%%%%%%%%%%%%%%%%%%%%%%%%%%%%%%
While most publicly available motion datasets, such as Human3.6M \cite{ionescu2013human3} and 3DPW \cite{von2018recovering}, are primarily designed for 3D pose estimation from 2D images, PROX \cite{von2018recovering} and LH26M \cite{li2022lidarcap} provide depth data from RGBD cameras and LiDAR sensors, respectively. However, the depth data in PROX are relatively noisy for sensor limitations and the point cloud in LH26M tends to be sparse for being captured from a far distance. In this work, we provide the MOVIN dataset with synchronized pairs of 3D point cloud and motion data, designed for full-body motion capture from 3D point cloud data.

\begin{figure}[t]
  \captionsetup{labelfont=bf,textfont=it}
  \centering
  \includegraphics[width=.8\linewidth]{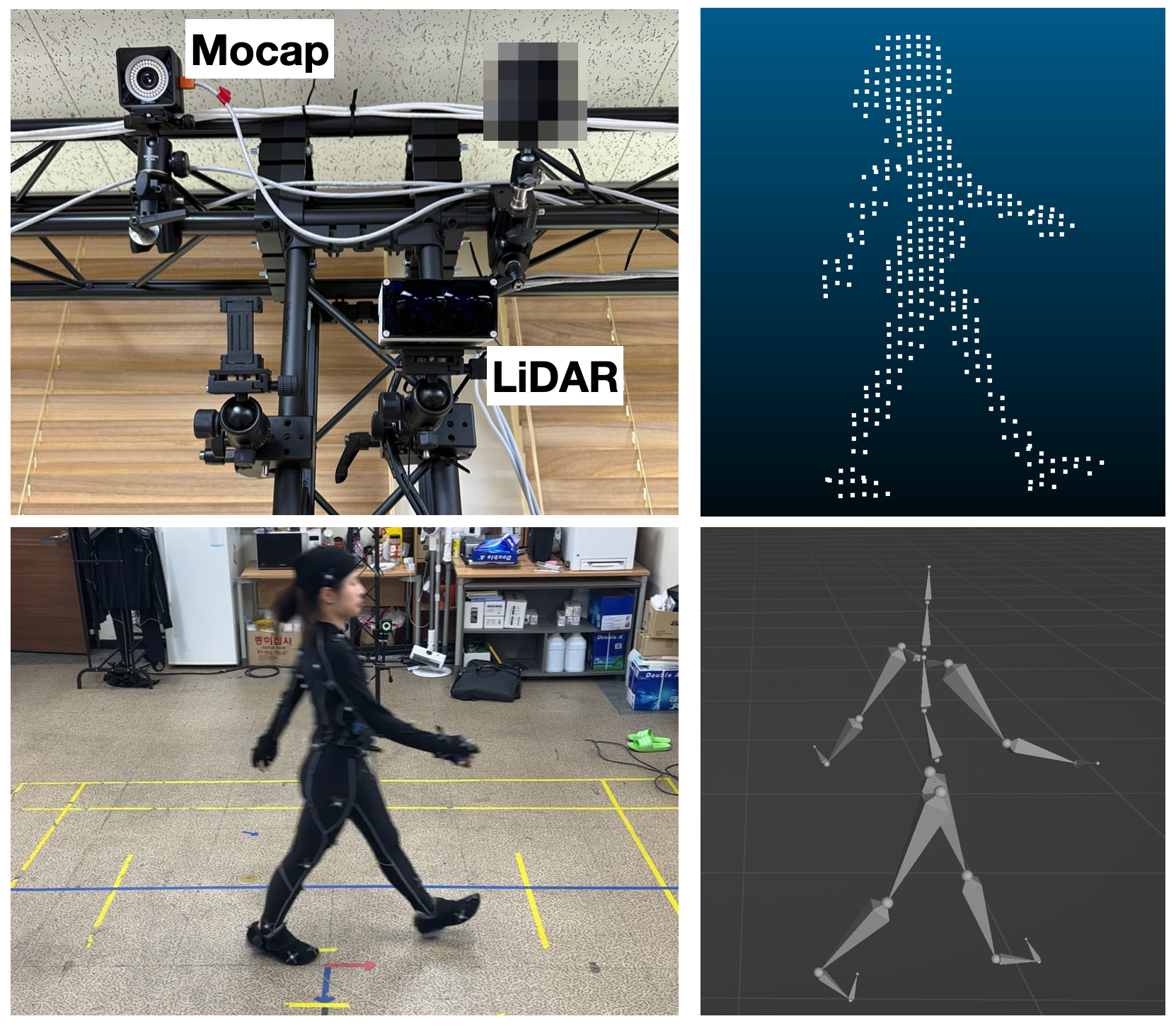}
  \caption{The integrated optical motion capture and LiDAR system, with representations of resulting point cloud data and skeletal motion capture.}
  \label{fig:mocap_system}
\end{figure}

\begin{table}[ht]
\captionsetup{labelfont=bf,textfont=it}
\small
\centering
\begin{tabular}{ccccc}
\hline
             & \multicolumn{2}{c}{Train set} & \multicolumn{2}{c}{Test set} \\ \hline
\# Subjects  & \multicolumn{2}{c}{8}         & \multicolumn{2}{c}{2}        \\
Motion type  & Static      & Locomotion      & Static      & Locomotion     \\
\# Frames    & 56,535      & 75,134          & 12,472      & 17,038         \\
Elapsed time$_{min}$ & 47          & 62              & 10          & 14     \\ \hline
&&&& \\
\end{tabular}

\begin{tabular}{p{4.3cm}p{3cm}} \hline
Static & Locomotion \\ \hline
T-pose, A-pose, Idle, Hands on waist & Walking \\
Elbows bent up, down & Jogging \\
Bow, Look, Roll head & Running \\
Windmill arms, Touch toes & Crouching \\
Twist torso, Hula hoop, Lean & Transitions \\
Lunge, Squat, Jumping Jack & Moving backward \\ 
Kick, Turn & Jumping \\
Walk / Run in place & Sitting on the floor \\
\hline
\end{tabular}
\caption{Dataset composition details and motion categories.}
\label{tab:datacomposition}
\end{table}

%%%%%%%%%%%%%%%%%%%%%%%%%%%%%%%%%%%%%%%%%%%%%%%%%%%%%%%%%%%%%%%%%%%%%%%%%%%%%%
\paragraph*{Motion capture System.}
%%%%%%%%%%%%%%%%%%%%%%%%%%%%%%%%%%%%%%%%%%%%%%%%%%%%%%%%%%%%%%%%%%%%%%%%%%%%%%

We employed the OptiTrack system \cite{optitrack-link} comprising 21 PRIME 13 high-speed infrared cameras to capture human motion. By tracking passive reflective markers positioned on the subject's body keypoints, the system accurately records joint positions and orientations. Using an optical-based motion capture system offers the advantage of avoiding global location errors that may arise during extended recording sessions, which is a common issue with IMU-based motion capture systems. Given that our Movin dataset requires prolonged collections of motion data, an optical-based motion capture system is highly suitable for this purpose.

%%%%%%%%%%%%%%%%%%%%%%%%%%%%%%%%%%%%%%%%%%%%%%%%%%%%%%%%%%%%%%%%%%%%%%%%%%%%%%
\paragraph*{LiDAR sensor.}
%%%%%%%%%%%%%%%%%%%%%%%%%%%%%%%%%%%%%%%%%%%%%%%%%%%%%%%%%%%%%%%%%%%%%%%%%%%%%%
LiDAR sensors emit laser light to accurately measure distances and generate high-resolution 3D maps of the surrounding environment and objects inside. The resulting 3D point cloud data provides detailed information about objects' geometries.

To capture 3D point cloud data of moving subjects, we utilized the ML-X model \cite{soslab2023} by SOSLAB, a high-performance solid-state LiDAR operating at a frequency of $20$ Hz. The ML-X model offers a wide field of view with a $120$ degree horizontal and $35$ degree vertical coverage, generating a detailed point cloud at a resolution of $56\times192$. This point cloud provides precise 3D global coordinates along with light intensity values. One notable advantage of this solid-state LiDAR system is capturing depth and intensity in both 3D and 2D image formats; this allows for treating the point cloud data similar to images in image-motion datasets. Moreover, the ML-X model exhibits minimal distortion compared to other spinning-type LiDAR systems, making it well-suited for capturing fast and complex movements with body overlaps.

%%%%%%%%%%%%%%%%%%%%%%%%%%%%%%%%%%%%%%%%%%%%%%%%%%%%%%%%%%%%%%%%%%%%%%%%%%%%%%
\paragraph*{Data acquisition.}
%%%%%%%%%%%%%%%%%%%%%%%%%%%%%%%%%%%%%%%%%%%%%%%%%%%%%%%%%%%%%%%%%%%%%%%%%%%%%%
Motion and point cloud data are captured simultaneously, as shown in Figure \ref{fig:mocap_system}. Subjects performed two main types of static movements and locomotion. Details on the dataset composition and motion categories are provided in Table \ref{tab:datacomposition}.

To extract only the points belonging to human subjects, we applied background filtering to the captured point cloud. The refined point cloud contains approximately $200$ to $300$ points per frame. We recorded only the 3D position data and excluded the intensity data, as it is not significant among subjects who wear identical black motion capture suits. Furthermore, such intensity data does not provide information about typical human clothing.

After aligning the captured motion and point cloud data to a shared global coordinate frame, we synchronized the time frames and downsampled the motion data to $20$Hz, which matches the operating frequency of the LiDAR sensor.

%%%%%%%%%%%%%%%%%%%%%%%%%%%%%%%%%%%%%%%%%%%%%%%%%%%%%%%%%%%%%%%%%%%%%%%%%%%%%%
\section{Input/Output representation}
\label{sec:data_representation}
%%%%%%%%%%%%%%%%%%%%%%%%%%%%%%%%%%%%%%%%%%%%%%%%%%%%%%%%%%%%%%%%%%%%%%%%%%%%%%
\begin{figure}[t]
  \captionsetup{labelfont=bf,textfont=it}
  \centering
  \includegraphics[width=.9\linewidth]{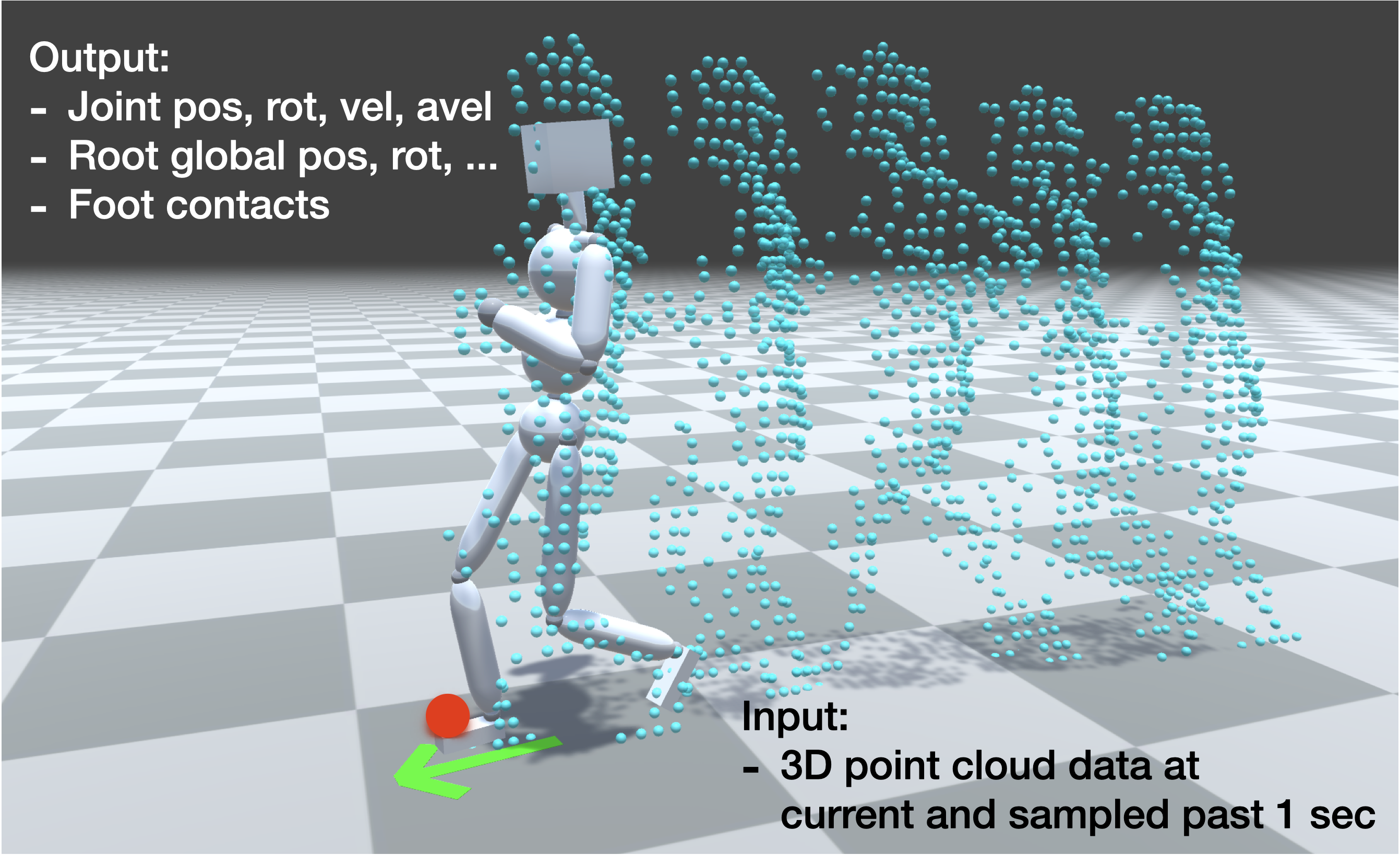}
  \caption{Illustration of the input and output in inference time. The gray skeleton represents the output joint positions, rotations, and velocities at the current time frame. Blue particles represent the 3D point cloud sampled in the current and the past 1-second time window. The green arrow and red sphere on the ground denote the global translation of the root and foot contacts, respectively.}
  \label{fig:input_output}
\end{figure}

Figure~\ref{fig:input_output} illustrates the input and output of our framework for a current frame $t$ in inference time. The input consists of the current 3D point cloud data as well as subsampled past 1-second point cloud data (at a frequency of 20 Hz). Each frame contains 256 3D points. We include four past point clouds sampled from 5, 10, 15, and 20 frames prior to the current frame. To ensure consistent input dimensions, we randomly discard points beyond 256 and perform zero-position padding for frames with fewer than 256 points. This results in the input $\vp_t = [p_t, p_{t-5}, p_{t-10}, p_{t-15}, p_{t-20}]$, where $p_t \in \mathbb{R}^{256 \times 3}$. Considering historical point cloud significantly improves the quality of results (Sec.~\ref{subsec:ablation}).

The output for a current frame $t$ consists of a global pose feature $\vg_t$ and a local pose feature $\vx_t$. The global pose feature includes the character's root position, rotation, velocity, angular velocity, and foot contacts, represented as $\vg_t = [r^l, r^r, \dot{r}^l, \dot{r}^r, c] \in \mathbb{R}^{17}$, where $r^l \in \mathbb{R}^3$, $r^r \in \mathbb{R}^6$, $\dot{r}^l \in \mathbb{R}^3$, $\dot{r}^r \in \mathbb{R}^3$, and $c \in \mathbb{R}^2$. The local pose feature consists of joint local positions, rotations, velocity, and angular velocity relative to the parent joint, denoted as $\vx_t = [x^l, x^r, \dot{x}^l, \dot{x}^r] \in \mathbb{R}^{n_j \times 15}$, where $x^l \in \mathbb{R}^{n_j \times 3}$, $x^r \in \mathbb{R}^{n_j \times 6}$, $\dot{x}^l \in \mathbb{R}^{n_j \times 3}$, and $\dot{x}^r \in \mathbb{R}^{n_j \times 3}$. Here, $n_j$ represents the number of joints.
%%%%%%%%%%%%%%%%%%%%%%%%%%%%%%%%%%%%%%%%%%%%%%%%%%%%%%%%%%%%%%%%%%%%%%%%%%%%%%
\section{MOVIN Framework}
\label{sec:method}
%%%%%%%%%%%%%%%%%%%%%%%%%%%%%%%%%%%%%%%%%%%%%%%%%%%%%%%%%%%%%%%%%%%%%%%%%%%%%%
\begin{figure*}[ht]
  \captionsetup{labelfont=bf,textfont=it}
  \centering
  \includegraphics[width=.85\textwidth]{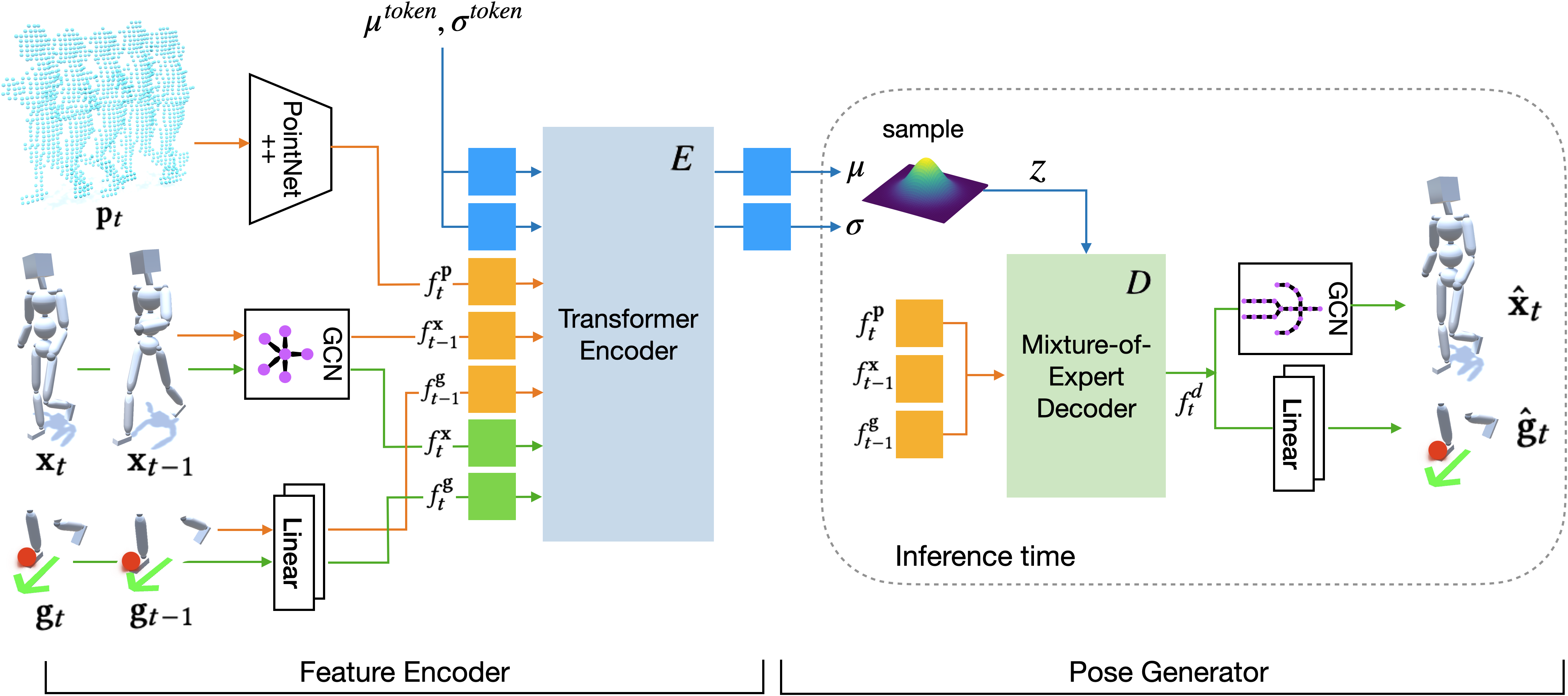}
  \caption{Overview of MOVIN framework. The model separates into the Feature Encoder and the Pose Generator. At inference time, only the Pose Generator and the embedding modules of Feature Encoder are used. Given the sampled point cloud sequence $\vp_t$, our model generates current global and local pose features $\hat{\vg}_t$, $\hat{\vx}_t$, which is used as a condition the next time frame. 
  }
  \label{fig:overview}
\end{figure*}

%We propose MOVIN, the first data-driven generative model for real-time 3D motion capture using LiDAR 3D point clouds. 
The MOVIN framework, illustrated in Fig.~\ref{fig:overview}, is based on an autoregressive conditional variational autoencoder (CVAE) architecture. During the training phase, MOVIN reconstructs the current global and local pose features, denoted as $\vg_{t}$ and $\vx_{t}$, respectively, using the 3D point cloud history $\vp_t$, as well as the previous global and local pose features $\vg_{t-1}$ and $\vx_{t-1}$. In addition, the model is trained to shape the latent variable $z$ into a Gaussian distribution. The framework comprises two components: the feature encoder, responsible for mapping input features to the latent distribution, and the pose generator, which generates global and local pose features.

During the inference phase, the embedding modules of the feature encoder and pose generator are used exclusively to generate the output pose features $\hat{\vx}_t$ and $\hat{\vg}_t$ for the current time step. The input consists of a randomly sampled latent variable $z$, while the conditions include the history of 3D point cloud $\vp_t$ and the output pose features from the previous frame, $\hat{\vx}_{t-1}$ and $\hat{\vg}_{t-1}$.

%%%%%%%%%%%%%%%%%%%%%%%%%%%%%%%%%%%%%%%%%%%%%%%%%%%%%%%%%%%%%%%%%%%%%%%%%%%%%%
\subsection{Feature Encoder}
\label{subsec:encoder}
%%%%%%%%%%%%%%%%%%%%%%%%%%%%%%%%%%%%%%%%%%%%%%%%%%%%%%%%%%%%%%%%%%%%%%%%%%%%%%
The feature encoder takes the previous output pose features $\vg_{t-1}$, $\vx_{t-1}$, the history of 3D point cloud data $\vp_t$, and the current target pose features $\vg_t$, $\vx_t$ as inputs and encodes them to a latent variable $z$ in a Gaussian distribution $\mathcal{N}(\mu, \sigma)$. The feature encoder is composed of embedding modules that individually embed the input features, and a transformer encoder that captures the relationships between the embedded vectors, allowing the model to learn a distribution of possible pose features for the current time frame.

\paragraph*{Embedding modules.}
\label{par:embedding}

To capture contextual information at different scales in the input 3D point cloud data, we utilize PointNet++~\cite{qi2017pointnet++} for extracting the embedded vector $f_t^{\vp}$ from the history of point cloud $\vp_{t}$:
\begin{equation} \label{eq:pointnet}
f_t^{\vp} = \text{PointNet++}(\vp_{t}) \in \mathbb{R}^{5 \times C},
\end{equation}
where $C$ denotes the number of channels.

For pose features, we employ separate embeddings for global and local pose features (as mentioned in Sec.\ref{sec:data_representation}). The Graph Convolution Network (GCN)\cite{yan2019convolutional, li2019actional, shi2019skeleton, jang2022motion} is used to reduce the spatial resolution of the input local features $\vx_t$ and $\vx_{t-1}$ while preserving the body part structure. Additionally, a two-layer MLP is utilized to embed the global pose features $\vg_t$ and $\vg_{t-1}$. The pose feature embedding process can be defined as follows:
\begin{equation} \label{eq:gcn_mlp}
f_t^{\vx} = \text{GCN}(\vx_{t}) \in \mathbb{R}^{C}, \quad\,\,
f_t^{\vg} = \text{MLP}(\vg_{t}) \in \mathbb{R}^{C}
\end{equation}
By applying the same procedure to $\vx_{t-1}$ and $\vg_{t-1}$, we obtain $f_{t-1}^{\vx}$ and $f_{t-1}^{\vg}$.

After feature embedding, we have five embedded vectors: $f^{\vp}_t$, $f^{\vx}_t$, $f^{\vx}_{t-1}$, $f^{\vg}_{t}$, and $f^{\vg}_{t-1}$. These vectors serve as input for the subsequent transformer encoder. Moreover, $[f_t^{\vp},\ f^{\vx}_{t-1},\ f^{\vg}_{t-1}]$ are used as conditions for the Pose Generator.

\paragraph*{Transformer encoder.}
To model the correlation between human joints and local clusters in the corresponding point cloud, we utilize a transformer architecture~\cite{vaswani2017attention}. The transformer encoder $E$ takes learnable tokens $[\mu^{token},\sigma^{token}]$ and concatenated embedded vectors $[f^{\vp}_t,\ f^{\vx}_{t-1},\ f^{\vg}_{t-1},\ f^{\vx}_{t},\ f^{\vg}_t]$ as inputs. These inputs are encoded to obtain the parameters of a Gaussian distribution $\mathcal{N}(\mu, \sigma)$. The reparameterization trick is then applied to transform these parameters and obtain the decoder input distribution $z \in \mathbb{R}^{C}$:
\begin{equation} \label{eq:trans_enc}
E(z|f^{\vp}_t,\ f^{\vx}_{t-1},\ f^{\vg}_{t-1},\ f^{\vx}_{t},\ f^{\vg}_t) = \mathcal{N}(z; , \mu, \sigma)
\end{equation}

%%%%%%%%%%%%%%%%%%%%%%%%%%%%%%%%%%%%%%%%%%%%%%%%%%%%%%%%%%%%%%%%%%%%%%%%%%%%%%
\subsection{Pose Generator}
\label{subsec:generator}
%%%%%%%%%%%%%%%%%%%%%%%%%%%%%%%%%%%%%%%%%%%%%%%%%%%%%%%%%%%%%%%%%%%%%%%%%%%%%%
Given the sampled latent variable $z$, the pose generator is an autoregressive model that generates current global and local pose features, $[\hat{\vx}_t,\ \hat{\vg}_t]$, conditioned on the embedded vectors of the sampled point cloud history and the previous pose features $[f_t^\vp,\ f^\vx_{t-1},\ f^\vg_{t-1}]$. 
Since a single LiDAR sensor often suffers from self-occlusions between body parts, it increases ambiguity between the obtained point cloud and the ground truth full-body pose. To address this, we sample a latent vector $z$ from a prior distribution and use the point cloud as the condition to generate plausible body motion.

Inspired by motionVAE~\cite{ling2020character}, we incorporate a Mixture-of-Expert (MoE) decoder, which we have observed empirically to enhance pose construction and reduce visual artifacts. MoE methods are often used to divide the problem space into distinct partitions assigned to a fixed number of neural network experts. A gating network is then employed to determine the relative contribution of each expert's prediction when computing the final output or prediction. In our framework, the MoE decoder generates an output, and this output is further expanded using expanding modules to obtain the final full-body pose and foot contacts. These expanding modules use inverse forms of the embedding modules found in the feature encoder.

\paragraph*{Mixture-of-Expert decoder}
The MoE decoder $D$ consists of eight expert networks with identical structures. A single shared gating network is incorporated to blend the weights of the experts, thereby determining the weights of the decoder network. Given the latent variable $z$ and the set of condition features $[f^{\vp}_t,\ f^{\vx}_{t-1},\ f^{\vg}_{t-1}]$, the MoE decoder $D$ computes the output $f^d_t$ as follows:
\begin{equation} \label{eq:decode}
f^d_t = D(z,\ f_t^\vp,\ f_{t-1}^\vx,\ f_{t-1}^\vg) \in \mathbb{R}^{2C},
\end{equation}
where $f_t^\vp$ represents the embedded vector of the current point cloud, and $f_{t-1}^\vx$, $f_{t-1}^\vg$ represent the embedded vectors of the previous pose features.

\paragraph*{Expanding modules.}
The output of the MoE decoder $f^d_t$ is further processed by De-GCN and De-MLP modules, which have architectures symmetric to the embedding modules. These modules expand the dimensions of the output to obtain the final global and local pose features $\hat{\vg}_t$ and $\hat{\vx}_t$ as follows:
\begin{equation} \label{eq:expanding}
\hat{\vx}_t = \text{De-GCN}(f^d_t\left[:C\right]), \quad\,\,
\hat{\vg}_t = \text{De-MLP}(f^d_t\left[C:\right])
\end{equation}

%%%%%%%%%%%%%%%%%%%%%%%%%%%%%%%%%%%%%%%%%%%%%%%%%%%%%%%%%%%%%%%%%%%%%%%%%%%%%%
\section{Training}
\label{sec:training}
%%%%%%%%%%%%%%%%%%%%%%%%%%%%%%%%%%%%%%%%%%%%%%%%%%%%%%%%%%%%%%%%%%%%%%%%%%%%%%
The overall model is trained by minimizing the reconstruction $\mathcal{L}_{rec}$ and KL-divergence $\mathcal{L}_{kl}$ losses. The reconstruction loss comprises both the local and global pose feature reconstruction losses. The local reconstruction loss quantifies the L1 errors in joint space with respect to the parent and character space with respect to the character’s root. Similarly, the global reconstruction loss measures the L1 errors between the generated and ground truth global root position, rotation, velocity, and foot contacts. In addition, the KL-divergence loss $\mathcal{L}_{kl}$ regularize distribution of latent variable $z$ to be near the prior distribution $\mathcal{N}(\boldsymbol{0}, \boldsymbol{I})$.

The total loss function is thus:
\begin{equation} \label{eq:rec}
\begin{aligned}
    \mathcal{L}_{total} = & \mathcal{L}_{rec} + w_{kl}\mathcal{L}_{kl} \\
    \mathcal{L}_{rec} = &\| \hat{\vx}_t - \vx_t \|_1 + \| FK(\hat{\vx}_t)- FK(\vx_t) \|_1 + \| \hat{\vg}_t - \vg_t \|_1,
    % \mathcal{L}_{rec} = & \mathcal{L}^{\vx}_{rec} + \mathcal{L}^{\vg}_{rec} \\
    % \mathcal{L}^{\vx}_{rec} = &\| \hat{\vx}_t - \vx_t \|_1 + \| FK(\hat{\vx}_t)- FK(\vx_t) \|_1 \\
    % \mathcal{L}^{\vg}_{rec} = &\| \hat{\vg}_t - \vg_t \|_1. 
\end{aligned}
\end{equation}
where the first and second terms of $\mathcal{L}_{rec}$ denote local reconstruction loss, and the last term is for global reconstruction loss. $w_{kl}$ is weight of KL-divergence loss.

\paragraph*{Implementation details.}
The AdamW optimizer was used over 120 epochs, with a learning rate of $10^{-4}$. Loss weight $w_{kl}$ was set as $1$. 
In the embedding module, the GCN layer comprises 1 spatial convolution layer along with body part pooling. The 2-layer MLP comprised a feed-forward network with 256 hidden units and ReLU activation. Meanwhile, PointNet++ is made up of 3 set abstraction layers.
Transformer encoder $E$ comprised 2 layers of 64 channels with 4 heads, and the MoE decoder $D$ consists of 8 identically structured expert networks and a single gating network. 
The gating network is also a 3-layer feed-forward network with 256 hidden units. 
Expanding module have architectures symmetric to the embedding modules.
To prevent the covariate shift during autoregressive inference, we set the prediction length as 8 frames for training. Scheduled sampling was also utilized in our model to enable long-term generation by making the model robust to its own errors. With four 12GB 2080ti GPUs, training took around 60 hours in an end-to-end manner.

%%%%%%%%%%%%%%%%%%%%%%%%%%%%%%%%%%%%%%%%%%%%%%%%%%%%%%%%%%%%%%%%%%%%%%%%%%%%%%
\section{Evaluation and Experiments}
\label{sec:Evaluation}

To validate the effectiveness of our method, we conducted comprehensive quantitative and qualitative evaluations against state-of-the-art methods. We selected VIBE \cite{kocabas2020vibe} and MotionBERT \cite{zhu2022motionbert} as representative baselines, which are vision-based approaches. To match the skeleton hierarchy for comparison, we applied BVH conversion and an optimization-based retargeting \cite{jin2018aura} to the output parameters of the baseline methods. Furthermore, we downsampled the retargeted outputs to $20$ fps to align with our output framerate. We disabled any postprocessing for all methods to ensure a fair comparison of the network architectures. For visual animation results, please refer to the supplementary video. 

Due to the unavailability of public datasets containing synchronized video, LiDAR point cloud, and motion capture data, our experiments were conducted solely on our held-out test set. The test set consists of two subjects with heights of 162 cm and 170 cm and each subject performed motion categories of static movement and locomotion. The length of the entire test set is around $25$ minutes.

Additionally, we performed ablation studies to investigate the impact of our design choices, including the utilization of point cloud history and the implementation of the autoregressive scheme.

The quantitative metrics include mean position error (M*PE), rotation error (M*RE), linear velocity error (M*LVE), and angular velocity error (M*AVE), for the Pelvis (P) and other body Joints (J). Joint position and linear velocity errors are calculated using forward kinematics in the pelvis coordinate frame. In addition for MOVIN, we assessed contact accuracy by comparing ground truth and predicted contact labels obtained by applying a threshold of $0.5$ to the predicted contact probabilities.

Lastly, we showcase a real-time motion capture demo on the wild unseen subject and discuss about effect of post-processing.

\subsection{Comparison with State-of-the-art methods}
\label{subsubsec:SOTA}
We conducted inference for VIBE and MotionBERT using their public code and note that these baselines are offline methods with fixed input sequence lengths of $16$ and $243$, respectively. In contrast, our model, MOVIN, performed per-frame prediction with a sliding window size of 1 to simulate real-time usage. We specifically measured pelvis errors for MOVIN, the errors significantly impact the quality of the output full-body pose. Since explicit global localization is not supported by the baselines, we did not measure pelvis errors for them and provided them with ground truth pelvis trajectory for qualitative analysis.

\begin{table}[t]
\captionsetup{labelfont=bf,textfont=it}
\small
\setlength{\tabcolsep}{4.6pt}
\begin{tabular}{cccccc}
\midrule
& MJPE${}_{cm}$ & MJRE${}^\circ$ & MJLVE${}_{cm}$ & MJAVE${}^\circ$ & Jitt. \\
\midrule
GT & $-$ & $-$ & $-$ & $-$ & $446.87$\\
VIBE & $10.86$ & $18.39$ & $2.39$ & $3.16$ & $1103.15$\\
MotionBERT & $\uline{10.62}$ & $\uline{18.05}$ & $\textbf{1.75}$ & $\textbf{2.24}$ & $\textbf{395.11}$\\
MOVIN & $\textbf{6.21}$ & $\textbf{10.12}$ & $\uline{1.89}$ & $\uline{2.75}$ & $\uline{871.53}$\\
\midrule
& MPPE${}_{cm}$ & MPRE${}^\circ$ & MPLVE${}_{cm}$ & MPAVE${}^\circ$ & Cont.$_{\%}$ \\
\midrule
MOVIN &  $4.42$ & $11.64$ & $2.46$ & $4.94$ & $94.28$\\
\midrule
\end{tabular}
\caption{Quantitative measures of MOVIN and state-of-the-art methods. Pelvis (P) errors are only measured for MOVIN  since the baselines cannot accurately capture global translation.}
\label{tab:quansota}
\end{table}

Table \ref{tab:quansota} presents the quantitative evaluation results of MOVIN and the state-of-the-art methods. MOVIN demonstrated a significant improvement over MotionBERT in terms of average joint position and rotation errors, with margins of approximately $4.41$ cm and $7.93$ degrees, respectively. However, for joint linear and angular velocities, MotionBERT exhibited slightly better performance. This advantage can be attributed to MotionBERT's utilization of a longer input window of $243$ frames, enabling it to maintain continuity and achieve smoother transitions in the output. Notably, the output motions of MotionBERT displayed lower jitter values compared to the ground truth, indicating an over-smoothing effect that can be clearly observed in the supplementary video. VIBE showed similar position and rotation errors to MotionBERT but suffered from severe jittering in the output poses, as indicated by the large jitter value in Table \ref{tab:quansota}.

\begin{figure*}[t]
    \captionsetup{labelfont=bf,textfont=it}
    \begin{subfigure}{\textwidth}
    \centering
    \includegraphics[width=\textwidth]{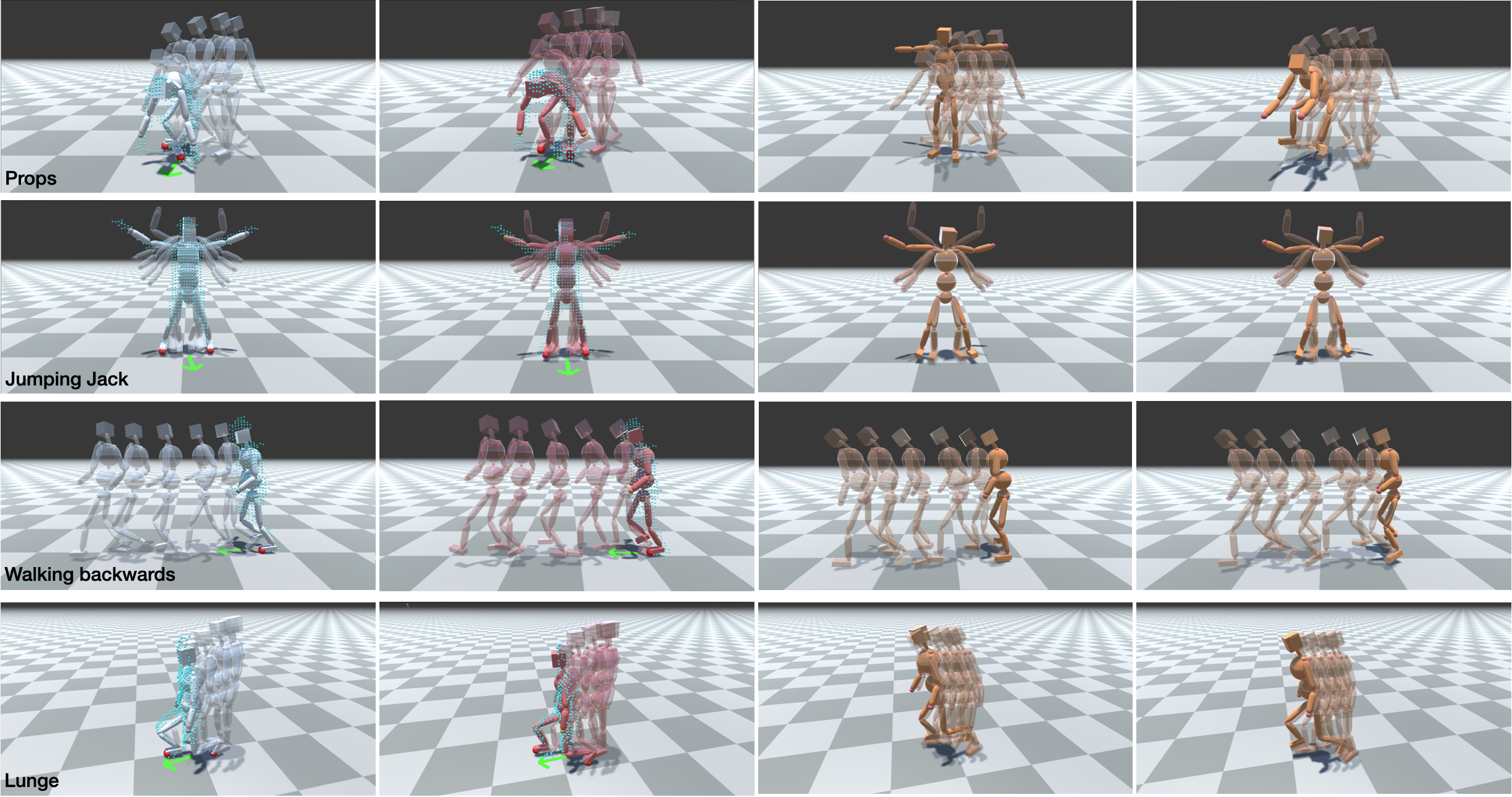}
    \caption{Male / 170 cm}
    \label{fig:qual_man}
    \end{subfigure}
    
    \begin{subfigure}{\textwidth}
    \centering
    \includegraphics[width=\textwidth]{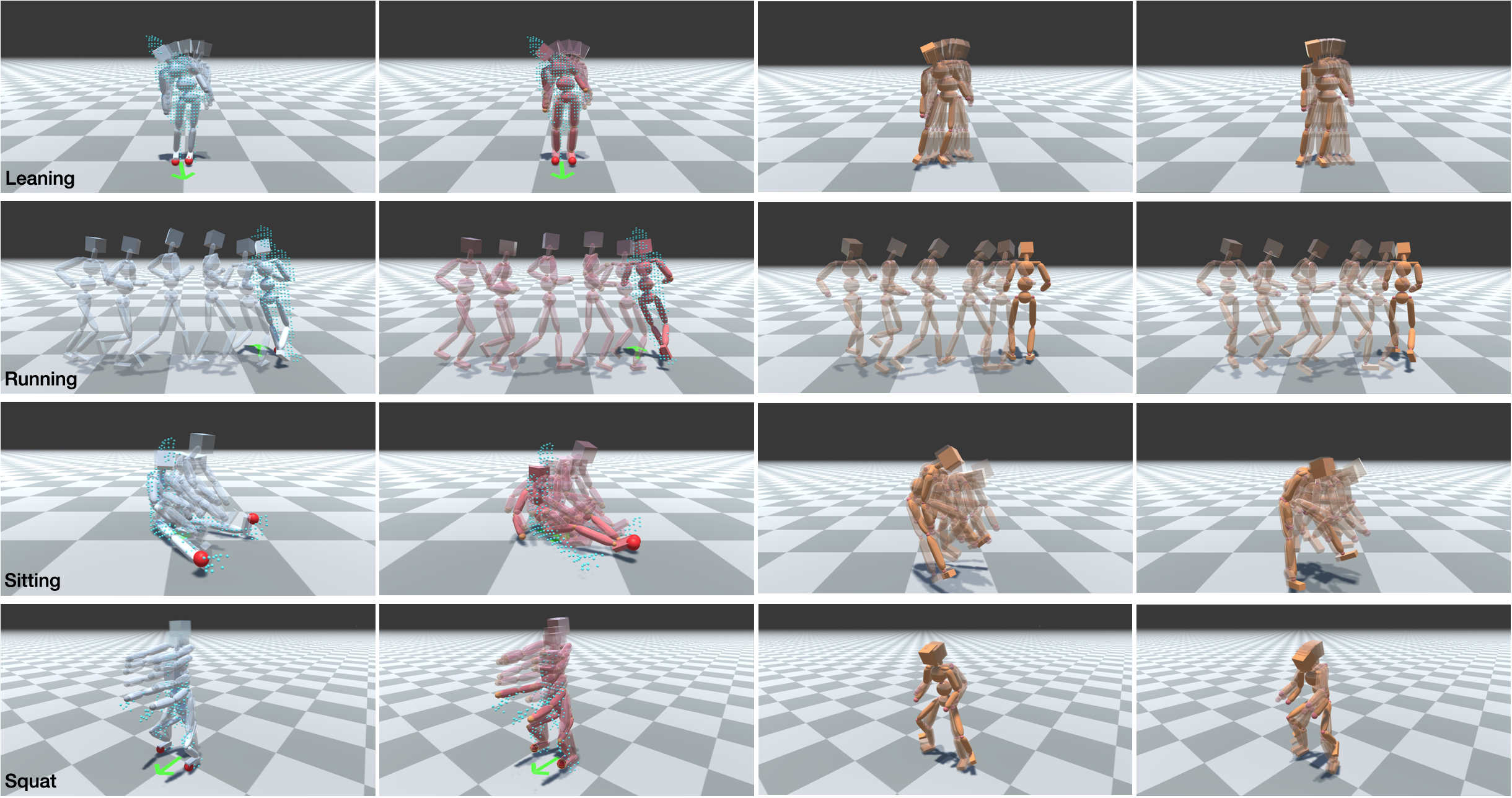}
    \caption{Female / 162 cm}
    \label{fig:qual_woman}
    \end{subfigure}
\caption{Qualitative comparisons of full-body motion outputs: Ground Truth, MOVIN-Ours, VIBE, and MotionBERT (from left to right). Our model, MOVIN, accurately generates output motion that closely resembles the ground truth, with natural joint trajectories. In contrast, baseline methods often suffer from issues such as oversmoothing, inaccurate pose, or temporal discontinuities with noticeable jitter.}
\label{fig:qualitative}
\end{figure*}

Figure \ref{fig:qualitative} presents two sets ($170$ cm male and $162$ cm female) of four-column images representing the ground truth (GT), and output full-body motions from MOVIN, VIBE, and MotionBERT, respectively. The robustness of MOVIN is evident as it generates plausible full-body motion, regardless of the subject body shape and across diverse action categories. In general, MOVIN preserved details in the ground truth and maintained temporal continuity in the output full-body motion. Please refer to our supplementary video for a comprehensive evaluation comparing MOVIN with the baselines.

Regarding the global localization performance, MOVIN exhibited average pelvis position and rotational errors of $4.42$ cm and $11.64$ degrees, respectively. The snapshots from our real-time application, depicted in Figure \ref{fig:realtime}, reveal that the output global trajectory is well aligned with that of the user. 

\subsection{Ablation study}
\label{subsec:ablation}

\begin{table}[t]
\captionsetup{labelfont=bf,textfont=it}
\small
\setlength{\tabcolsep}{4.6pt}
\begin{tabular}{cccccc}
\midrule
& MJPE${}_{cm}$ & MJRE${}^\circ$ & MJLVE${}_{cm}$ & MJAVE${}^\circ$ & Jitt. \\
\midrule
GT & $-$ & $-$ & $-$ & $-$ & $446.87$\\
w/o past pcd & $6.70$ & $11.49$ & $1.95$ & $2.76$ & $919.15$\\
w/ past poses & $7.25$ & $12.67$ & $\textbf{1.80}$ & $2.44$ & $\textbf{708.08}$\\
w/o autoreg. & $\textbf{6.09}$ & $\textbf{9.71}$ & $\uline{2.07}$ & $3.15$ & $1118.68$\\
512 points & $6.39$ & $10.15$ & $1.91$ & $\textbf{2.74}$ & $929.60$\\
MOVIN & $\uline{6.21}$ & $\uline{10.12}$ & $\uline{1.89}$ & $\uline{2.75}$ & $\uline{871.53}$ \\
\midrule
& MPPE${}_{cm}$ & MPRE${}^\circ$ & MPLVE${}_{cm}$ & MPAVE${}^\circ$ & Cont.$_{\%}$ \\
\midrule
w/o past pcd & $4.98$ & $12.03$ & $1.67$ & $6.76$ & $92.79$\\
w/ past poses & $5.44$ & $12.34$ & $1.58$ & $5.68$ & $93.93$\\
w/o autoreg. & $4.45$ & $\textbf{11.39}$ & $\textbf{1.48}$ & $\uline{5.83}$ & $\textbf{94.43}$\\
512 points & $\textbf{4.35}$ & $11.83$ & $1.51$ & $5.90$ & $\uline{94.38}$\\
MOVIN &  $\uline{4.42}$ & $\uline{11.64}$ & $\uline{1.50}$ & $\textbf{4.94}$ & $94.28$\\
\midrule
\end{tabular}
\caption{
Quantitative measures of MOVIN and ablation models. The term "w/o past pcd" denotes the variant trained without incorporating point clouds sampled from a previous time window. "w/ past poses" refers to the version that includes past poses. "w/o autoreg." signifies the variant that employs a non-autoregressive pose generator. Lastly, "512 points" designates the model variant that utilizes a point cloud consisting of 512 points.
}
\label{tab:quanablation}
\end{table}

The aim of our ablation study is to validate our design choices of utilizing point cloud history as input, excluding past poses from input, selecting an optimal number of input points, and implementing autoregression in both training and inference phases. Table \ref{tab:quanablation} presents quantitative metrics for five different ablation models: one without past point cloud input, one with past poses as input, one with 512 points as input, one without autoregression, and our proposed model, MOVIN.

\begin{figure}[t]
  \captionsetup{labelfont=bf,textfont=it}
  \centering
  \includegraphics[width=\linewidth]{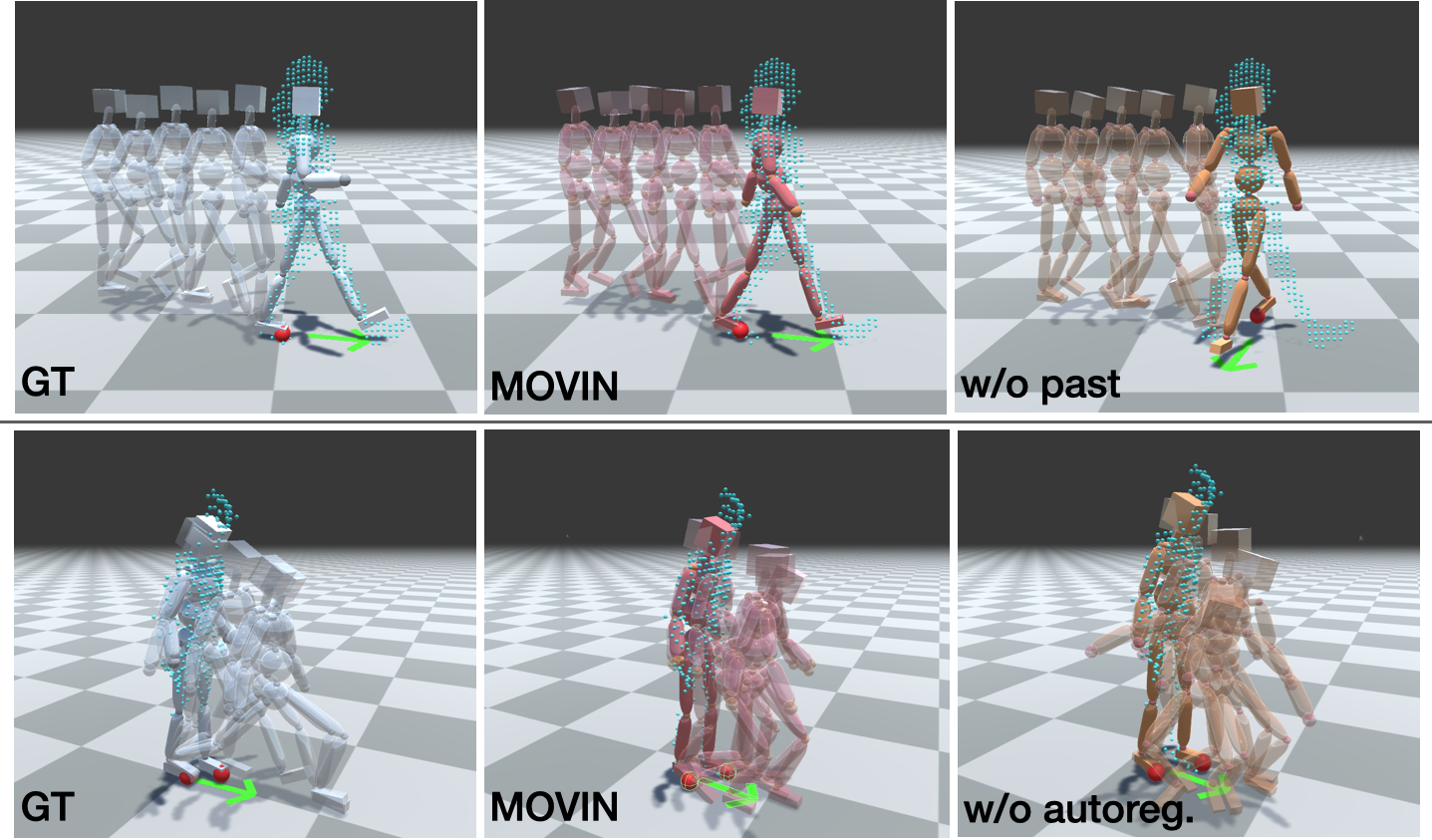}
  \caption{Visual comparisons of ablation models. Without past point clouds as input, the model exhibits abrupt changes in the global heading direction and incorrect movement. The model with a non-autoregressive Pose Generator produces outputs with unrelated poses between frames.}
  \label{fig:ablation}
\end{figure}

\paragraph*{Past point cloud sequence and poses.}
The model without a past point cloud sequence underperformed compared to the proposed model. it showed an increase of $0.5$ cm in average joint position error and $1.4$ degrees in rotation error. Specifically for the pelvis joint, the average position error and angular velocity error increased by $0.5$ cm and $1.8$ degrees, respectively. In the output motion sequence, we observed that this model often fails to maintain a temporal continuity, especially for rapid movements or cases when certain body parts are occluded by others (i.e. walking sideways or sitting) as shown in Figure~\ref{fig:ablation} (1st row); this results in abrupt changes in the global heading direction and body poses.

Providing past poses performed the worst among the methods. We hypothesize that simply providing previously generated poses leads to the model suffering from accumulated errors in the autoregressive input, thereby making it challenging to recover from inaccurate predictions.

\paragraph*{Autoregression.}
The model without autoregression shows no significant differences compared to the proposed model in terms of position and rotation errors. However, there are noticeable increases in linear and angular velocity errors for the joints, particularly for the pelvis where the angular velocity error rises by approximately $0.9$ degrees. Additionally, the jitter value increases by around $250$. These findings suggest that incorporating autoregression and exposing the model to accumulated prediction errors during training, enables it to robustly handle such errors during inference and produce stable and continuous poses in the output sequence. The result in Figure~\ref{fig:ablation} (2nd row) shows that the ablated model produces the discontinuous poses during the Lunge.

\paragraph*{Number of the input points.}
Doubling the number of input points (from 256 to 512) increases computation time proportionally but does not bring a significant improvement in performance. To achieve real-time inference at the pace of a 20Hz LiDAR sensor, we opted for 256 points as input.

%%%%%%%%%%%%%%%%%%%%%%%%%%%%%%%%%%%%%%%%%%%%%%%%%%%%%%%%%%%%%%%%%%%%%%%%%%%%%%
\subsection{Real-time Motion Capture Demo}
\label{sec:realtime}
%%%%%%%%%%%%%%%%%%%%%%%%%%%%%%%%%%%%%%%%%%%%%%%%%%%%%%%%%%%%%%%%%%%%%%%%%%%%%%
\begin{figure*}[ht]
  \captionsetup{labelfont=bf,textfont=it}
  \centering
  \includegraphics[width=\textwidth]{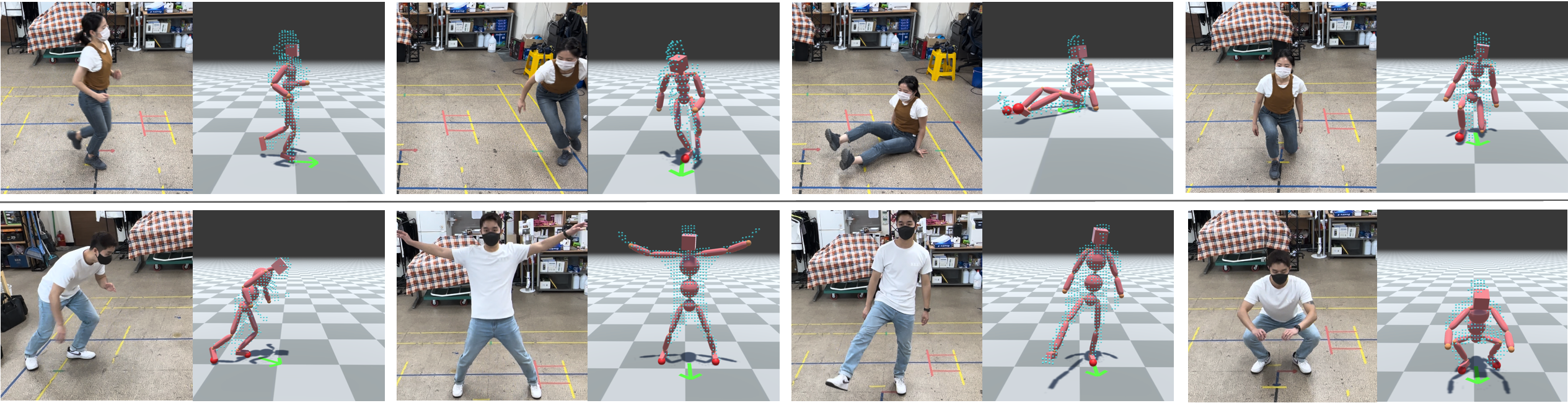}
  \caption{Real-time motion capture results. For each user's pose, the left shows reference images and the right shows the corresponding captured full-body pose. Our framework robustly captures both body dimensions and a wide range of motions.}
  \label{fig:realtime}
\end{figure*}

Figure \ref{fig:realtime} showcases snapshots from our real-time motion capture system with a single LiDAR, implemented in Unity3D. Compared to recent methods that use multiple RGB cameras, our system does not require offline calibration and captures the subject's motion in real-time, allowing users to check the results immediately. In the first row, a female subject with a height of 159cm performs challenging motions such as sitting on the floor and lunging, which are accurately captured by our system. The second row highlights the application's ability to capture dynamic actions from a male subject with a height of 175cm, including running, jumping jacks, kicking, and squats. Our framework not only accurately tracks these diverse movements but also effectively captures the subjects' body dimensions. Please refer to our supplementary video for a detailed demonstration of our model's real-time performance.

%%%%%%%%%%%%%%%%%%%%%%%%%%%%%%%%%%%%%%%%%%%%%%%%%%%%%%%%%%%%%%%%%%%%%%%%%%%%%%
\subsection{Effect of Post-processing}
\label{subsec:post-processin}
%%%%%%%%%%%%%%%%%%%%%%%%%%%%%%%%%%%%%%%%%%%%%%%%%%%%%%%%%%%%%%%%%%%%%%%%%%%%%%
As the real-time pose generation method cannot consider future poses, the output motion may exhibit foot sliding. To address this issue, we employ predicted contact labels and utilize inverse kinematics to correct foot positions. The target foot position is determined by interpolating between the previous and the output foot positions. Our supplemental video demonstrates the impact of this post-processing by comparing output motions with and without it.
%%%%%%%%%%%%%%%%%%%%%%%%%%%%%%%%%%%%%%%%%%%%%%%%%%%%%%%%%%%%%%%%%%%%%%%%%%%%%%
\section{Limitations and Future Work}
\label{sec:limitation}
%%%%%%%%%%%%%%%%%%%%%%%%%%%%%%%%%%%%%%%%%%%%%%%%%%%%%%%%%%%%%%%%%%%%%%%%%%%%%%
While our proposed model, MOVIN, successfully captures diverse motions in real-time, it is important to acknowledge the existing limitations for future research.

\begin{figure}[t]
  \captionsetup{labelfont=bf,textfont=it}
  \centering
  \includegraphics[width=.8\linewidth]{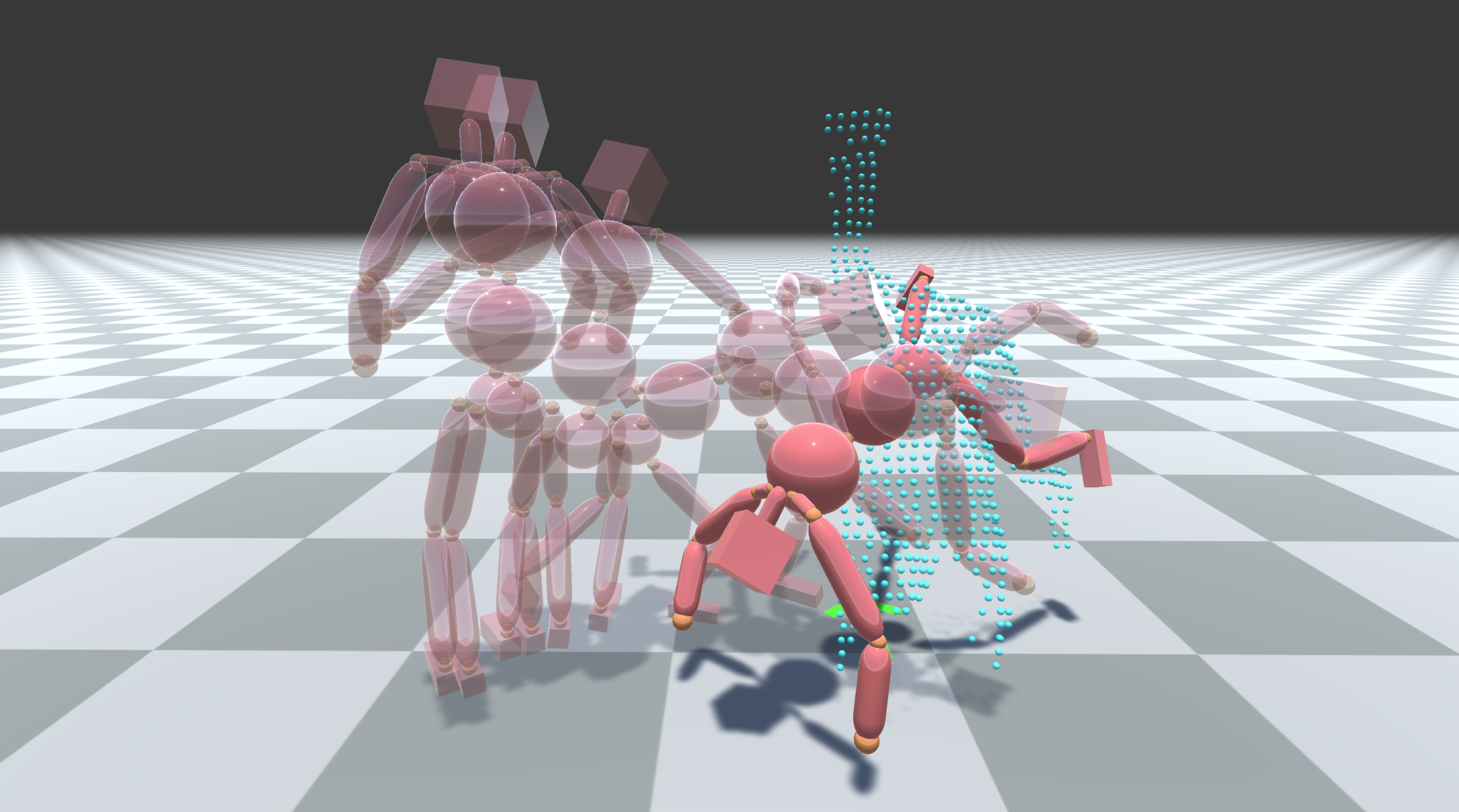}
  \caption{An example of failure case. When the user performs an unseen motion (windmill), the model generates a pose that does not align with the corresponding point cloud.}
  \label{fig:failure}
\end{figure}

One limitation of our model arises when encountering unseen motions, as demonstrated by the windmill motion example shown in Figure \ref{fig:failure}. In these cases, the generated pose does not align with the corresponding point cloud, highlighting the need for improved generalization to novel or uncommon movements. Expanding the size of the training dataset by incorporating a wider range of motion variations is a potential approach to address this limitation.

Another is the relatively low frames-per-second (fps) performance of our current implementation, primarily due to the operating frequency of the LiDAR sensor. To overcome this, future work could explore techniques such as point cloud upsampling or hardware improvements to enhance the fps rate. Upsampling the point cloud data can provide denser and more frequent input information, resulting in smoother output motion and a higher frame rate.

Additionally, while our model demonstrates reasonable handling of self-occlusions between body parts, it struggles in cases of severe occlusions, such as when a subject curls up or the environment is cluttered with objects. To improve performance in such scenarios, considering the use of multiple LiDAR sensors positioned from different angles could be a viable solution. By capturing aligned point clouds from multiple perspectives, the model can access more detailed information and enhance its capture performance.
%%%%%%%%%%%%%%%%%%%%%%%%%%%%%%%%%%%%%%%%%%%%%%%%%%%%%%%%%%%%%%%%%%%%%%%%%%%%%%
\section{Conclusion}
\label{sec:conclusion}
%%%%%%%%%%%%%%%%%%%%%%%%%%%%%%%%%%%%%%%%%%%%%%%%%%%%%%%%%%%%%%%%%%%%%%%%%%%%%%

We present MOVIN, the first data-driven generative model for real-time full-body motion capture using a single LiDAR sensor. Our approach addresses the challenges of full-body tracking by eliminating the need for body-worn suits and devices while maintaining high-quality motion capture. MOVIN utilizes an autoregressive CVAE model to learn the distribution of pose variations from 3D point cloud data. By separately embedding global and local pose features, our model effectively learns the pose prior and accurately predicts the performer's 3D global information and local joint details. The proposed autoregressive Mixture-of-Expert decoder ensures temporal coherence across frames, resulting in natural and realistic motion. Our real-time application showcases MOVIN's robustness to accurately capture diverse motions from subjects with varying body shapes, demonstrating its effectiveness in real-world scenarios.

%%%%%%%%%%%%%%%%%%%%%%%%%%%%%%%%%%%%%%%%%%%%%%%%%%%%%%%%%%%%%%%%%%%%%%%%%%%%%%
\section*{Acknowledgement}
%%%%%%%%%%%%%%%%%%%%%%%%%%%%%%%%%%%%%%%%%%%%%%%%%%%%%%%%%%%%%%%%%%%%%%%%%%%%%%
This work was supported by IITP, MSIT, Korea (2022-0-00566) and NRF, Korea (2022R1A4A5033689).

% Qualitative figure
% qualitative comparison  figure.

%-------------------------------------------------------------------------
% bibtex
\bibliographystyle{eg-alpha-doi} 
\bibliography{reference}       

% biblatex with biber
% \printbibliography

\end{document}